\documentclass[a4paper]{VisionStyle}
\usepackage{epsfig}

\begin{document}

\title{Soft X-ray Spectroscopy of \object{NGC~1068} with {\em XMM-Newton} {\small{RGS}} and {\em Chandra} {\small{LETGS}}}

\author{A.\,Kinkhabwala\inst{1}, M.\,Sako\inst{1,2,3}\, E.\,Behar\inst{1}, S.M.\,Kahn\inst{1}, F.\,Paerels\inst{1}, A.C.\,Brinkman\inst{4}, J.S.\,Kaastra\inst{4}, R.L.J.\,van der Meer\inst{4}, M.F.\,Gu\inst{3,5}, D.A.\,Liedahl\inst{6}} 

\institute{Columbia Astrophysics Laboratory, 
	      Columbia University, 
	      550 West 120th Street, 
	      New York, NY 10027
\and
  Theoretical Astrophysics and Space Radiation Laboratory,
		California Institute of Technology,
		MC 130-33,
		Pasadena, CA 91125
\and
  Chandra Fellow
\and
  Space Research Organization of the Netherlands, 
              Sorbonnelaan 2, 3548 CA, 
              Utrecht, The Netherlands
\and 
  Center for Space Research,
	      Massachusetts Institute of Technology,
		Cambridge, MA, 02139
\and 
  Physics Department, 
              Lawrence Livermore National Laboratory,
              P.O. Box 808, L-41, 
              Livermore, CA 94550}

\maketitle 

\begin{abstract}

We present high-resolution soft-X-ray spectra of the prototypical 
Seyfert~2 galaxy, \object{NGC~1068}, taken with the  \emph{XMM-Newton}
Reflection Grating Spectrometer ({\small RGS}) and the \emph{Chandra} 
Low Energy Transmission Grating Spectrometer ({\small LETGS}).  Its rich 
emission-line spectrum is dominated by recombination in 
a warm plasma (bright, narrow radiative recombination continua provide the 
``smoking gun''), which is photoionized by the inferred nuclear power-law 
continuum.  Radiative decay following photoexcitation of resonant transitions 
also provides an important contribution.  A 
self-consistent model of a photoionized and photoexcited cone of gas 
is capable of reproducing the hydrogenic/heliumlike ionic line series 
in detail.  The ratio of photoexcitation to photoionization in the cone 
provides important geometric information such as the radial ionic column 
densities, which are consistent with absorption measurements (the warm 
absorber) in Seyfert~1 galaxies.  This strongly suggests that the emission 
spectrum we observe from \object{NGC~1068} emanates from its warm absorber.
The observed extent of the ionization-cone/warm-absorber in 
\object{NGC~1068} of about 300~pc implies that a large fraction of the 
gas associated with generic warm absorbers may typically
exist on the hundreds-of-parsec scale rather than much closer to the nucleus 
(e.~g., less than a parsec).  Spatially-resolved spectroscopy using the 
{\small LETGS} of two distinct 
emission regions yields two noticeably different spectra.  We show 
that these differences are solely due to differing radial column densities.  
A fairly flat distribution in 
ionization parameter $\xi=L_X/n_{\mathrm{e}}r^2$ (over at least 
$\log{\xi}\sim0$--3) is necessary to explain the inferred radial ionic column 
densities of all spectra.  We show that this must primarily be due to a 
broad density distribution 
$f(n_{\mathrm{e}})\propto n_{\mathrm{e}}^{-1}$ at each radius, spanning 
roughly $n_{\mathrm{e}}\sim0.1$--$100$~cm$^{-3}$.  Additional 
contributions to the soft-X-ray emission from hot, collisionally-ionized gas,
 if present, make a negligible contribution to the spectrum.

\end{abstract}

\section{Introduction}

In the unified model of active galactic nuclei ({\small AGN}), the 
observational properties of a particular {\small{AGN}} are determined simply 
by its orientation 
(\cite{akinkhabwala-C2:millerantonucci,akinkhabwala-C2:antonuccimiller}).  
For Seyfert~1 galaxies, we directly observe the intrinsic nuclear continuum, 
partially absorbed by photoionized outflowing material (warm absorber). 
We observe Seyfert~2 galaxies at an angle nearly perpendicular to the 
Seyfert~1 orientation.
From this vantage, the intrinsic nuclear continuum is highly obscured
(by the ``dusty torus''), and the spectrum is dominated by reprocessed 
emission from outlying clouds filling an ionization cone, which may be 
related to the warm absorber.

The first high-resolution soft-X-ray spectrum obtained of a Seyfert~2 galaxy 
was the \emph{Chandra} {\small{HETGS}} of Markarian~3 
(\cite{akinkhabwala-C2:mkn3}).  \cite*{akinkhabwala-C2:mkn3}
interpreted the bulk of this spectrum as due to recombination/radiative 
cascade following photoionization.  Radiative decay following photoexcitation 
is also required to explain the enhanced resonance lines in the heliumlike 
triplets.  This is consistent with outlying clouds irradiated by the
inferred nuclear continuum, as predicted in the unified model 
\cite*{akinkhabwala-C2:krolikkriss}.

We present below the first high-resolution X-ray spectrum of the 
X-ray-brightest, prototypical Seyfert~2 galaxy, \object{NGC~1068}, which was 
obtained with \emph{XMM-Newton} {\small RGS} (\cite{akinkhabwala-C2:rgs}).  
Using a simple model of an irradiated gas cone, we are able to find an
excellent fit to the detailed soft X-ray spectrum of \object{NGC~1068} (for 
all hydrogenic/heliumlike ions).  From the similarity of our inferred radial 
column densities to directly-measured absorption column densities (warm 
absorber) in Seyfert~1 galaxies, we argue that the observed 
hundreds-of-parsec-scale ionization cone of \object{NGC~1068} is identical to 
the warm absorber in this {\small AGN}.  A subsequent spectrum obtained with 
the \emph{Chandra} {\small LETGS} (\cite{akinkhabwala-C2:letgs}) confirms the 
{\small RGS} results, but, more importantly, allows for spatially-resolved 
spectroscopy of this extended source.

\section{Full \small{RGS/LETGS} Spectra of \object{NGC~1068}}
\label{akinkhabwala-C2_sec:full}

\begin{figure*}[!ht]
  \begin{center}
    \begin{minipage}[t]{3.5in}
      \vspace{-.1in}\hspace{0in}
        \centerline{\psfig{file=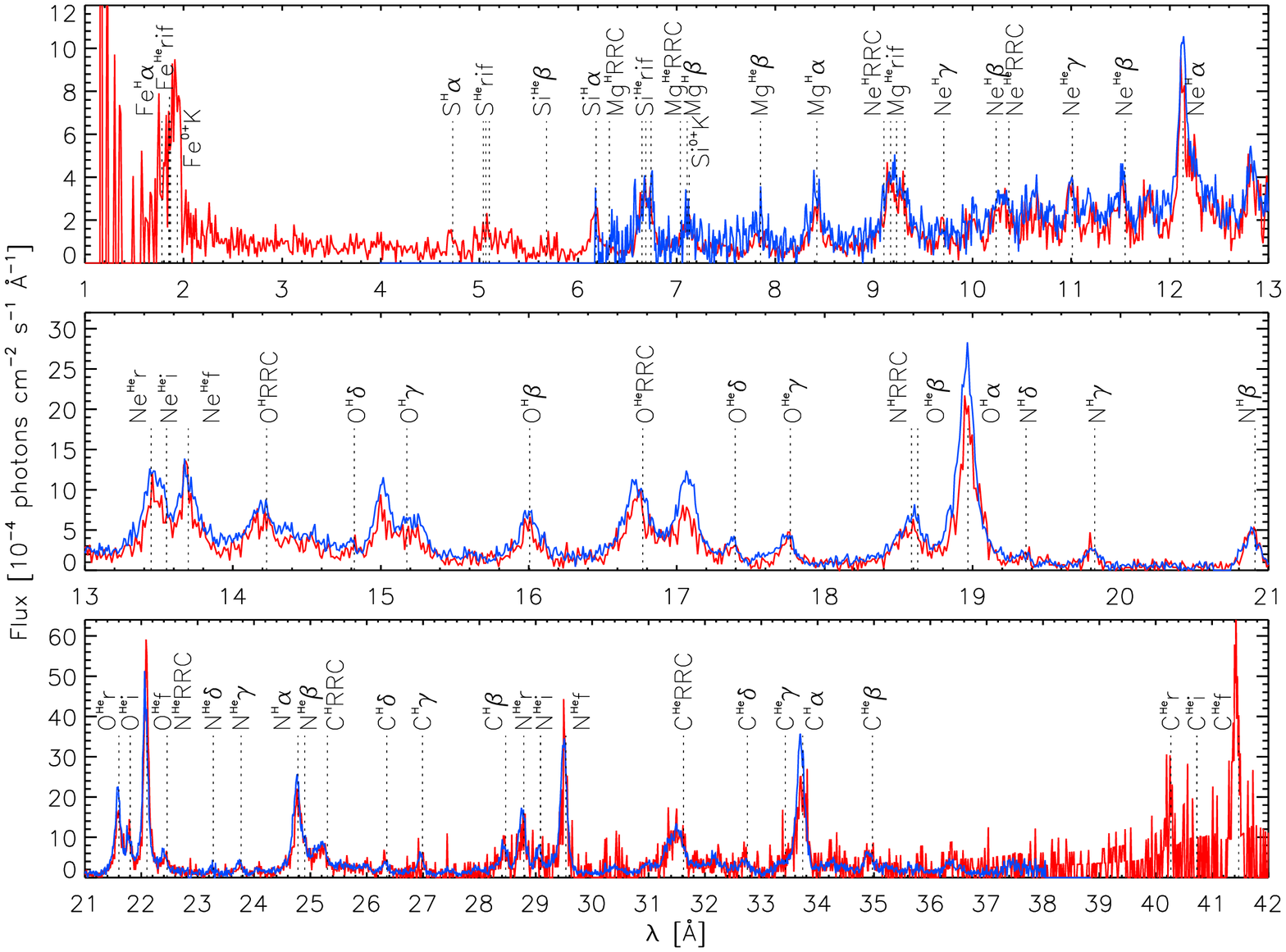, width=3.7in, angle=90}}
      \vspace{-.2in}
    \end{minipage}
  \end{center}
\caption{The RGS1/RGS2 (m=-1 order) (blue) and LETGS 
(combined m=-1,+1 
orders) (red) spectra of \object{NGC~1068}.  
Note the overall flux and wavelength agreement between both 
intruments.  The spectrum has been shifted to the \object{NGC~1068} rest frame 
($z=0.00379$).  The superscripts {\rm 'H'} (hydrogenic) and {\rm 'He'} 
(heliumlike) refer to the final-state ion (e.~g., 
{\rm 'C$^{\mathrm{He}}$RRC'} refers to recombination forming heliumlike 
carbon.)  Unlabelled features at $\lambda<18$~\AA\ are all due to Fe L-shell
transitions, whereas unlabelled features at $\lambda>18$~\AA\ are due to 
mid-Z-element (mostly S and Si) L-shell transitions.\label{akinkhabwala-C2_fig:full}}
\end{figure*}

Line emission dominates the {\small RGS}/{\small LETGS} soft-X-ray spectra 
of \object{NGC~1068} shown in Fig.~\ref{akinkhabwala-C2_fig:full} 
(for ease of comparison, all 
spectral plots appear at the end of the paper).
Emission lines from hydrogenic and heliumlike C, N, O, Ne, Mg, Si,
and S are all clearly detected.   Fe~L-shell emission lines from 
{\small Fe~XVII} to {\small Fe~XXIV} are present as well (unlabelled), 
with numerous transitions scattered between 9~\AA\ and 18~\AA.  Many 
higher-order resonant 
transitions (1s$\rightarrow$np) in the hydrogenic and heliumlike ions labelled 
$\beta$--$\delta$ are prominent, with evidence for 
strong emission from even higher order transitions as well.  
Several unidentified features at longer wavelengths
(e.~g., at 27.92, 30.4, 34.0--34.6, and 36.38~\AA) are 
likely due to L-shell emission from mid-Z ions such as sulfur.  The fluorescent
lines of neutral Fe (Fe$^{+0}$) and Si (Si$^{+0}$) are clearly detected in the 
{\small LETGS} spectrum.  We see no significant 
continuum emission in the spectrum.

All lines appear broader than expected for a monochromatic source and several
lines show evidence for significant blue shifts.  These are
due to intrinsic velocity distributions (hundreds of km/s) and are not 
instrumental effects.

The spectrum also includes very prominent radiative recombination 
continua ({\small{RRC}}) 
for hydrogenic and heliumlike C, N, and O, which
are produced when electrons recombine directly to the ground state in
these highly-ionized ions.  
{\small{RRC}} are smeared out for hot collisionally-ionized 
gas, but are narrow, prominent features for cooler photoionized gas.  The 
narrow width of these {\small{RRC}}
provide a direct measure of the recombining electron temperature 
(\cite{akinkhabwala-C2:liedahlpaerels,akinkhabwala-C2:liedahl}), which for 
\object{NGC~1068} is $kT\sim2-10$~eV.

\section{Model of Irradiated Gas Cone}
\label{akinkhabwala-C2_sec:cmd}

\begin{figure}[!ht]
  \begin{center}
    \begin{minipage}[t]{3.5in}
      \vspace{-.2in}\hspace{0in}
        \centerline{\psfig{file=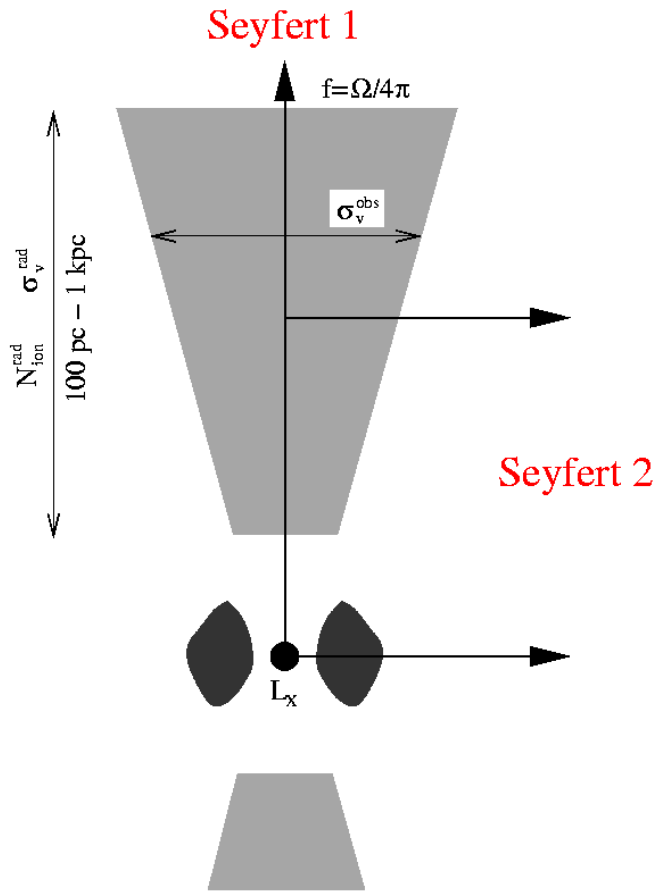, width=3.5in}}
      \vspace{-.2in}
    \end{minipage}	
  \end{center}
\caption{Schematic of our unified, warm-absorber/ionization-cone model 
for Seyfert~1 and 2 galaxies (not to scale).  The labels ``Seyfert 1'' and 
``Seyfert 2'' indicate the directions of the observer in the two cases. 
\label{akinkhabwala-C2_fig:cone}}
\end{figure}

We have constructed a fully self-consistent model of a gas cone (warm 
absorber) irradiated by a continuum source 
(Fig.~\ref{akinkhabwala-C2_fig:cone}).  
The nuclear region comprising the 
black-hole/ accretion-disc/comptonized-halo system is 
depicted as the central 
black circle.  Obscuration by the ``dusty torus'' is shown (in cross section) 
as two clouds on either side of the nucleus.  The parameters which comprise 
our two inferred global model parameters, the covering factor times nuclear 
luminosity $fL_X$ and the radial velocity width $\sigma^{\mathrm{rad}}_v$, 
are indicated, as well as the individual radial ionic column densities 
$N^{\mathrm{rad}}_{\mathrm{ion}}$.  $L_X$ is the total luminosity in a 
power-law with reasonable values for the index of $\Gamma=-1.7$ and 
energy range of 13.6~eV--100~keV.  

\begin{figure}[!ht]
  \begin{center}
    \begin{minipage}[t]{3.5in}
      \vspace{-.5in}\hspace{.15in}
        \centerline{\psfig{file=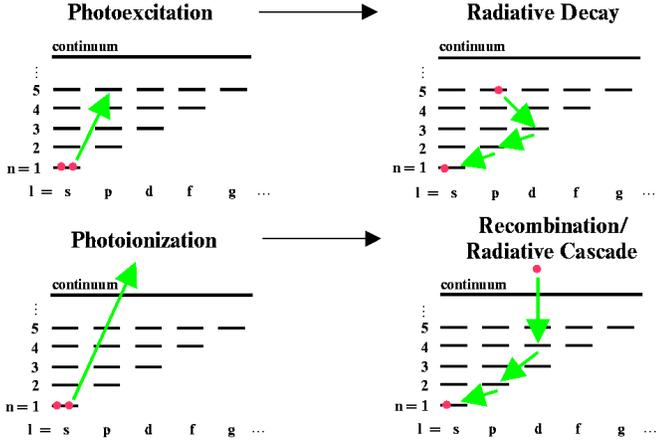, width=3.in, angle=-90}}
      \vspace{-.1in}
    \end{minipage}
  \end{center}
\caption{Simplified Grotrian diagram of the relevant atomic processes in 
hydrogenic/heliumlike ions.  Photoexcitations and photoionizations appear as 
absorption features in the Seyfert~1 view (left), but as line emission through
the inverse processes of radiative decay and recombination/radiative cascade
in the Seyfert~2 view (right).\label{akinkhabwala-C2_fig:grotrian}}
\end{figure}

Photoionization and photoexcitation in the ionization cone produce absorption 
features (warm absorber) in the Seyfert~1 view, but in the Seyfert~2 view, 
the inverse processes of recombination/radiative cascade and radiative decay, 
respectively, produce line emission (Fig.~\ref{akinkhabwala-C2_fig:grotrian}).
We use the new atomic code {\small FAC} (\cite{akinkhabwala-C2:fac}) 
to calculate the relevant atomic data for hydrogenic/heliumlike ions.  For 
ease of fitting, we have incorporated our model into
{\small XSPEC} (\cite{akinkhabwala-C2:xspec}) as a local model (``photo'').  
A complete discussion 
of the atomic calculations and astrophysical assumptions underlying our model 
is presented in \cite*{akinkhabwala-C2:model}.

For the specific case of \object{NGC~1068}, the intrinsic continuum is 
likely to be completely obscured, contributing 
no flux to the soft X-ray regime.  Also, the {\small NE} cone is 
much brighter than its counterpart in the {\small SW} 
(\cite{akinkhabwala-C2:young}), therefore, the covering factor $f$ 
applies to the {\small NE} cone alone (hence the asymmetry of
Fig.~\ref{akinkhabwala-C2_fig:cone}).

\subsection{Column Density}
\label{akinkhabwala-C2_sec:ex}

\begin{figure}[!ht]
  \begin{center}
    \begin{minipage}[t]{3.5in}
      \vspace{0in}\hspace{0in}
        \centerline{\psfig{file=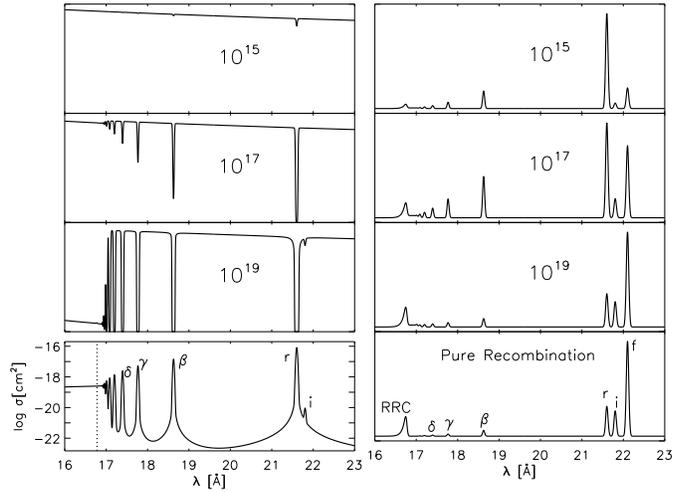, width=2.6in, angle=90}}
      \vspace{-.2in}
    \end{minipage}
  \end{center}
\caption{Effect of differing radial column densities on the reemitted 
spectrum for {\small{O~VII}} with the Seyfert~1 view in the top three panels
on the left and the Seyfert~2 view in the corresponding top three panels on 
the right.  The bottom left panel shows ionic cross section with
the photoexcitation/photoionization boundary indicated.  The bottom right panel
shows the spectrum expected for pure recombination.    All 
spectral axes are linear, but have arbitrary normalization.  Throughout,
we take a radial gaussian distribution with 
$\sigma^{\mathrm{rad}}_v=200$~km/s, transverse velocity distribution with 
$\sigma^{\mathrm{obs}}_v=400$~km/s, and temperature $kT=3$~eV.\label{akinkhabwala-C2_fig:opticaldepth}}
\end{figure}

We show the effect of varying the radial ionic column density on {\small{O~VII}} 
in Fig.~\ref{akinkhabwala-C2_fig:opticaldepth}.  The top three panels on the 
left show the radial ``Seyfert~1'' view through the outflow and down to the 
nucleus for radial column densities in {\small{O~VII}}  of $10^{15},10^{17},$ 
and $10^{19}$~cm$^{-2}$.
The corresponding top three panels on the right show the 
``Seyfert~2'' view roughly perpendicular to the axis of outflow and from which 
the nucleus is completely obscured (Fig.~\ref{akinkhabwala-C2_fig:cone}).  
All photons absorbed out of the power law in the ``Seyfert~1'' spectrum are 
reprocessed and reemitted to generate the ``Seyfert~2'' spectrum.
Radiative decay following photoexcitation dominates the ``Seyfert~2'' 
spectrum at low column
densities, whereas recombination/radiative cascade following photoionization 
dominates at high column densities (for comparison, pure recombination is 
shown in the lower right panel).  For hydrogenic ions, the behavior is 
similar (omitting the intercombination/forbidden lines).

\subsection{Velocity Width}
\label{akinkhabwala-C2_sec:sigmav}

\begin{figure}[!ht]
  \begin{center}
    \begin{minipage}[t]{3.5in}
      \vspace{-.23in}
        \centerline{\psfig{file=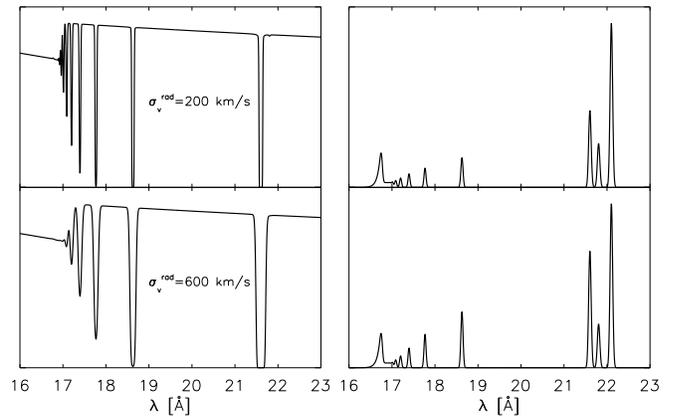, width=2.7in, angle=90}}
      \vspace{-.23in}
    \end{minipage}
  \end{center}
\caption{The effect of different $\sigma_v^{\mathrm{rad}}$ 
for the {\small{O~VII}}  spectrum using 
$N_{\mathrm{ion}}^{\mathrm{rad}}=10^{18}$~cm$^{-2}$, with the Seyfert~1 view 
on the left and the Seyfert~2 view on the right.  Spectra on the 
right-hand-side were convolved with the same transverse velocity distribution 
$\sigma_v^{\mathrm{obs}}=400$~km/s; we also assume $kT=3$~eV.
\label{akinkhabwala-C2_fig:sigmav}}
\end{figure}

In Fig.~\ref{akinkhabwala-C2_fig:sigmav}, we show the effect of varying the 
radial velocity width for {\small{O~VII}}.
Larger $\sigma_v^{\mathrm{rad}}$ enhance the importance of photoexcitation 
relative to photoionization.  

\section{Model Fit to the {\small RGS} Spectrum}

\begin{figure*}[!ht]
  \begin{center}
    \begin{minipage}[t]{3.5in}
      \vspace{-.1in}\hspace{0in}
        \centerline{\psfig{file=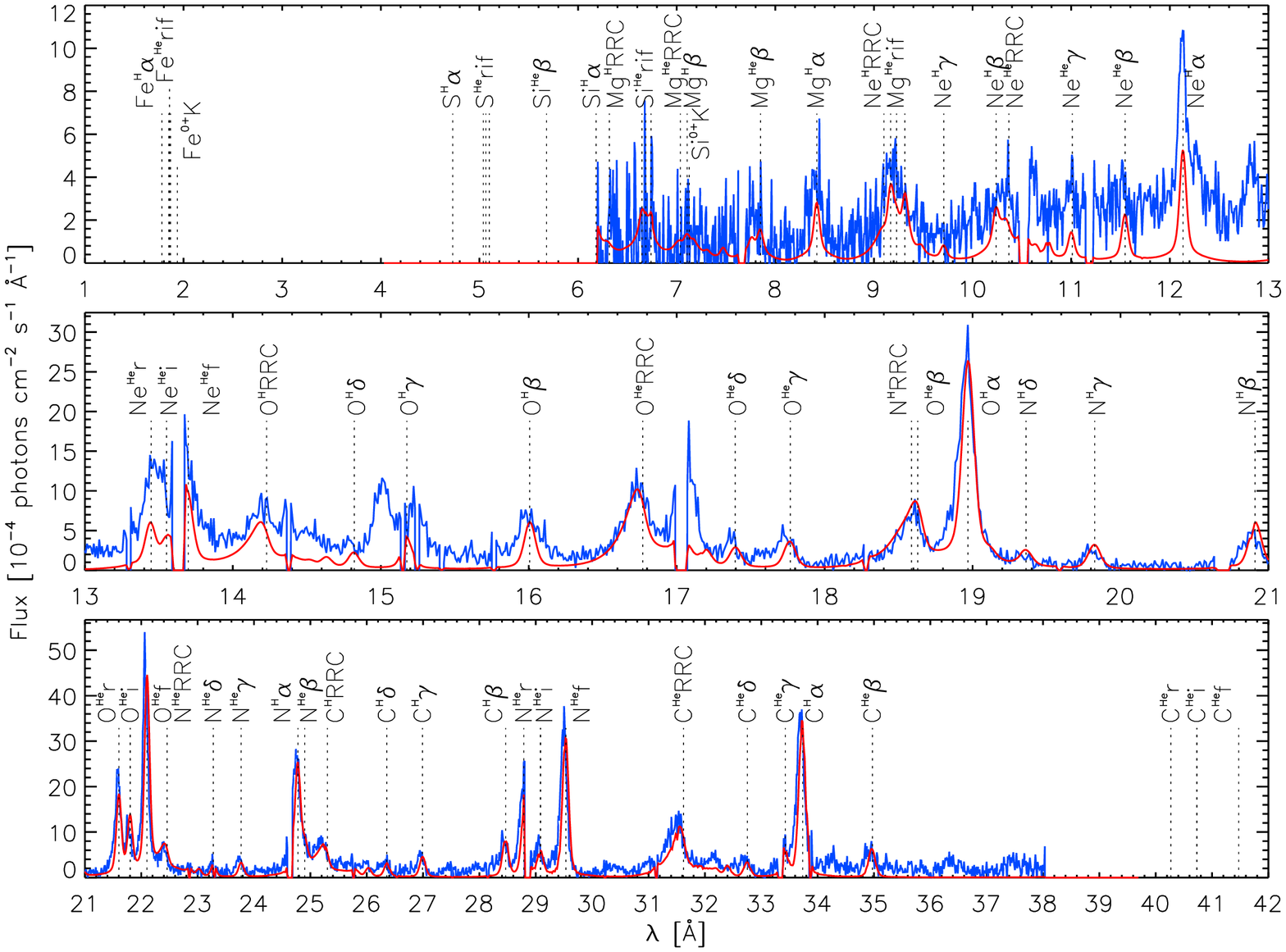, width=3.7in, angle=90}}
      \vspace{-.2in}
    \end{minipage}
  \end{center}
\caption{RGS1 spectrum (blue) and corresponding hydrogenic/heliumlike ion
fit (red) using parameters listed in Table~\ref{akinkhabwala-C2_tab:table}.  
The spectrum has been shifted to the \object{NGC~1068} rest frame 
($z=0.00379$).  Model wavelengths are those expected for no excess velocity 
shifts.  We employ the same labelling convention as in 
Fig.~\ref{akinkhabwala-C2_fig:full}.\label{akinkhabwala-C2_fig:rgs}}
\end{figure*}

We present our fit to the {\small RGS} spectrum of \object{NGC~1068} in 
Fig.~\ref{akinkhabwala-C2_fig:rgs}.  The parameters used for the fit are 
given in the first column of Table~\ref{akinkhabwala-C2_tab:table}.  
To convert observed flux to luminosity for the spectra of \object{NGC~1068}, 
we assume a distance of 14.4 Mpc 
(\cite{akinkhabwala-C2:bland-hawthorn}).  The column density of neutral 
hydrogen in our galaxy is taken to be $N_H=1$e$20$~cm$^{-2}$ (using the 
high-quality soft-X-ray spectrum itself), which is somewhat
lower than the standard column density to this source 
(\cite{akinkhabwala-C2:rgs}).  
For all fits in this paper, we assume recombining electron temperatures of 
$kT=2.5$~eV for {\small C~V}, $kT=3$~eV for {\small N~VI}, $kT=4$~eV for 
{\small C~VI, N~VII, O~VIII, and O~VIII}, and $kT=10$~eV for all other ions 
(which have poorly-determined {\small RRC}).  

\begin{figure}[!ht]
  \begin{center}
    \begin{minipage}[t]{3.5in}
      \vspace{-.24in}\hspace{0in}
        \centerline{\psfig{file=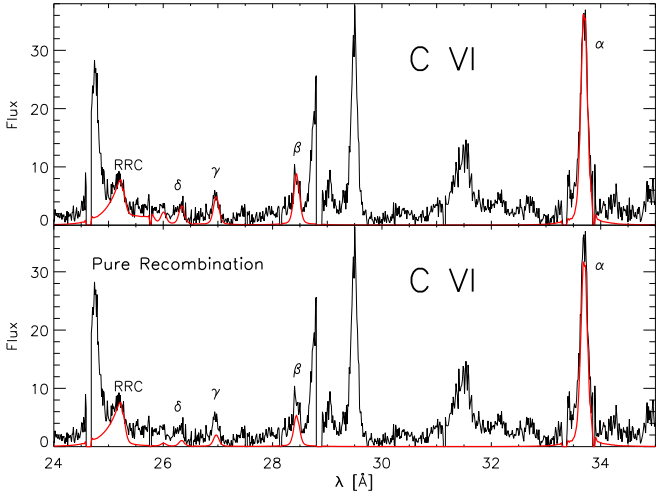, width=2.9in, angle=90}}     
      \vspace{-.25in}
    \end{minipage}
  \end{center}
\caption{The final fit to {\small C~VI} including recombination/radiative 
cascade following photoionization and radiative decay following 
photoexcitation (top).  Recombination alone (bottom) is unable to 
explain the excess emission in all resonant lines np$\rightarrow$1s.\label{akinkhabwala-C2_fig:C_fit}}
\end{figure}

\begin{figure}[!ht]
  \begin{center}
    \begin{minipage}[t]{3.5in}
      \vspace{-.24in}\hspace{0in}
        \centerline{\psfig{file=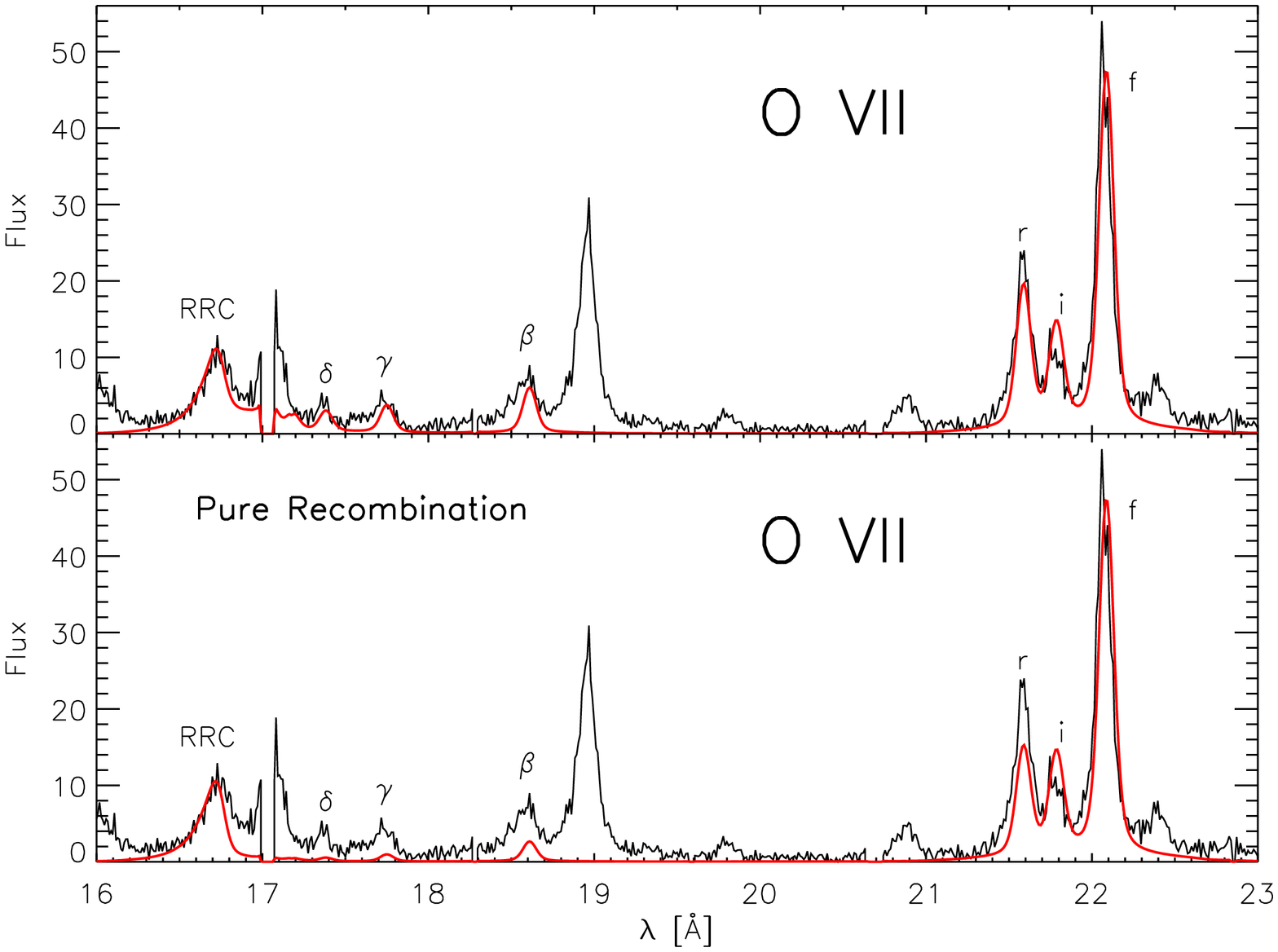, width=2.9in, angle=90}}
      \vspace{-.25in}
    \end{minipage}
  \end{center}
\caption{The final fit to {\small{O~VII}}  including recombination/radiative 
cascade following photoionization and radiative decay following 
photoexcitation (top).  Recombination alone (bottom) is unable to 
explain the excess emission in all resonant lines np$\rightarrow$1s.\label{akinkhabwala-C2_fig:O_fit}}
\end{figure}

\begin{figure}[!ht]
  \begin{center}
    \begin{minipage}[t]{3.5in}
      \vspace{0in}\hspace{-.12in}
        \centerline{\psfig{file=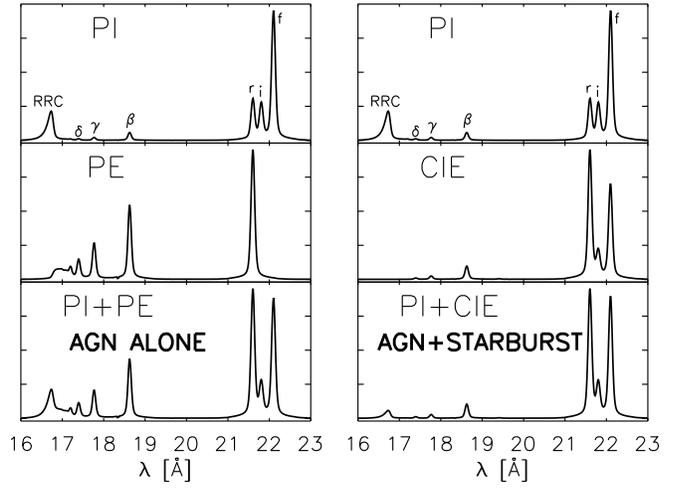, width=2.6in, angle=90}}
      \vspace{-.1in}
    \end{minipage}
  \end{center}
\caption{We demonstrate how to distinguish between hot collisional gas 
(e.~g., starburst region) and photoexcitation.  Starting with pure 
recombination following photoionization (``{\small PI}'' - top two panels, $kT=4$~eV), 
we self-consistently add radiative decay following 
photoexcitation (``PE''- middle left panel) assuming reasonable ionization 
cone parameters, or an additional hot gas component 
in collisional ionization equilibrium 
(``CIE'' - middle right panel, $kT=150$~eV) to obtain the bottom two panels.  
Note that both bottom panels have similar triplet ratios, implying that using
the triplet alone is insufficient to discriminate between these two scenarios.
However, the ``AGN ALONE'' panel has significantly stronger 
higher-order-series transitions (including the RRC) than the 
``AGN+STARBURST'' panel, demonstrating the diagnostic importance of these 
transitions.  (Normalization in each panel is arbitrary.)
\label{akinkhabwala-C2_fig:PIPECIE}}
\end{figure}

Velocity broadening of all model emission lines by 
$\sigma_v^{\mathrm{obs}}=400$~km/s  is necessary to fit the 
observed lines.

We illustrate the relative contributions of photoionization and photoexcitation
to the ionic line series for {\small C~VI} and {\small O~VII} in 
Figs.~\ref{akinkhabwala-C2_fig:C_fit} and \ref{akinkhabwala-C2_fig:O_fit}, 
respectively.  Pure recombination is unable to explain the anomalous strength 
of the higher-order-series transitions.  However, the self-consistent addition 
of photoexcitation allows for an excellent overall fit.  An additional 
collisional gas component {\em instead} of photoexcitation would be 
insufficient to explain the higher-order series transitions 
(Fig.~\ref{akinkhabwala-C2_fig:PIPECIE}).

\begin{table}[bht]
  \caption{Fit parameters for Figs.~\ref{akinkhabwala-C2_fig:rgs}, \ref{akinkhabwala-C2_fig:primary}, and \ref{akinkhabwala-C2_fig:secondary}.  {\rm 'H'} and {\rm 'He'} indicate hydrogenic and heliumlike, respectively.}
  \label{akinkhabwala-C2_tab:table}
  \begin{center}
    \leavevmode
    \footnotesize
    \begin{tabular}[!h]{llrcc}
      \hline \\[-5pt]
         & &  {\bf \small{RGS}} & {\bf Primary} & {\bf Secondary}\\
\hline
      \multicolumn{2}{l}{$fL_X$ [ergs/s]}    & 1e43 	  &	1e43	&	1e43	 \\
      \multicolumn{2}{l}{$\sigma_v^{\mathrm{rad}}$ [km/s]}    & 100	&60 	& 100	\\
      \hline 
      Ion & H/He & \multicolumn{3}{c}{Column Densities $N_{\mathrm{ion}}^{\mathrm{rad}}$ [cm$^{-2}$]} \\
      \hline 
      C~V    &He  & 5e17   & 4.5e17   & 2e17  \\
      C~VI   &H   & 7e17   & 6e17     & 1.5e17\\
      N~VI   &He  & 4e17   & 2.5e17   & 8e16  \\
      N~VII  &H   & 6e17   & 4e17     & 8e16  \\
      O~VII  &He  & 9e17   & 7e17     & 1.5e17\\
      O~VIII &H   & 1e18   & 6e17     & 2.5e17\\
      Ne~IX  &He  & 3e17   & 2.5e17   & 6e16  \\
      Ne~X   &H   & 2.5e17 & 1.5e17   & 6e16  \\
      Mg~XI  &He  & 2e17   & 1.2e17   & 1e16  \\
      Mg~XII &H   & 2e17   & 8e16     & 1e16  \\
      Si~XIII&He  & 2e17   & 2e17     & 2e16  \\
      Si~XIV &H   & 2e17   & 1.8e17   & 2e16  \\
      S~XV   &He  & ---    & 6e16     & 1e16  \\
      S~XVI  &H   & ---    & 9e16     & 1e16  \\
      \hline \\
      \end{tabular}
  \end{center}
\end{table}

\begin{figure}[!ht]
  \begin{center}
    \begin{minipage}[t]{3.5in}
      \vspace{-.2in}\hspace{0in}
        \centerline{\psfig{file=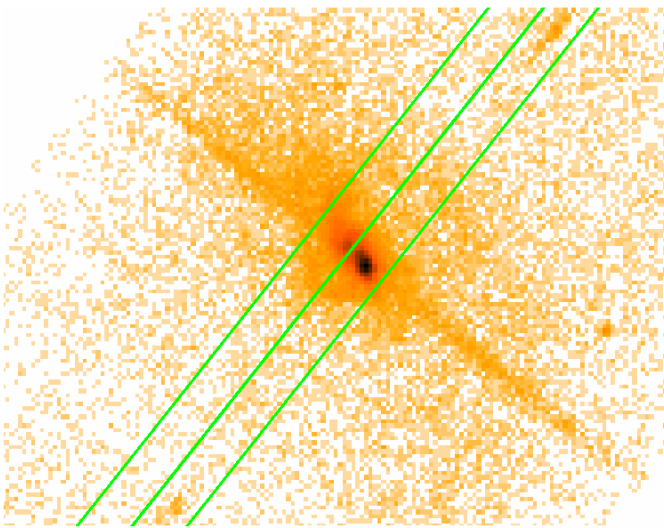, width=3.5in}}
      \vspace{-.2in}
    \end{minipage}
  \end{center}
\caption{Zero-order LETGS image (logarithmic) of \object{NGC~1068} 
oriented with N up and E to the left.  There are 
two general regions which we denote as the ``primary'' (brightest spot) and 
``secondary'' 
(accompanying NE spot).  The dispersion axis is parallel to the green lines,
which demarcate the cross dispersion regions we used to generate the
``primary'' and ``secondary'' spectra.  The apparent line of emission 
in the cross-dispersion direction and centered on the source
is instrumental (due to {\small{CCD}} readout).  The high-energy edge of
the dispersed spectrum can be seen in the upper right and lower left of the 
image.\label{akinkhabwala-C2_fig:image}}
\end{figure}

\begin{figure}[!ht]
  \begin{center}
    \begin{minipage}[t]{3.5in}
      \vspace{-.1in}\hspace{0in}
        \centerline{\psfig{file=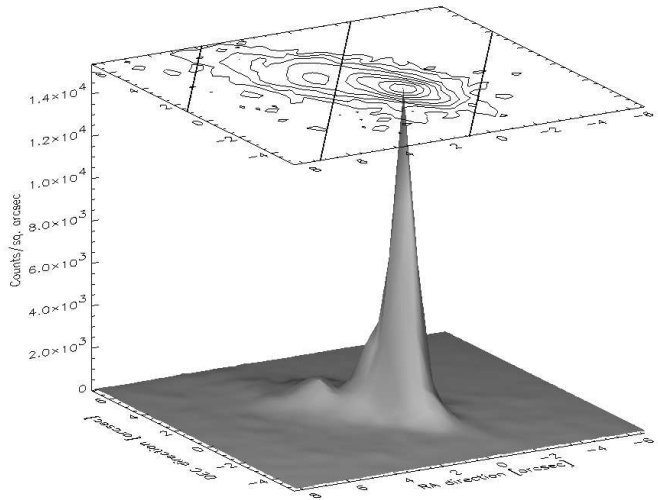, width=2.8in, angle=90}}
      \vspace{-.3in}
    \end{minipage}
  \end{center}
\caption{Zero-order LETGS image of \object{NGC~1068}.  The
relative brightness of the ``primary'' and ``secondary'' spots are clearly 
shown.  The straight lines in the ``factors-of-two'' contour image  are the 
same as the green lines in 
Fig.~\ref{akinkhabwala-C2_fig:image}, denoting the extraction regions for
the ``primary'' and ``secondary'' spectra.\label{akinkhabwala-C2_fig:3dimage}}
\end{figure}

\section{{\small{LETGS}} Spatially-Resolved Spectroscopy}

With the {\small{LETGS}}, it is possible to perform spatially-resolved 
spectroscopy by making cuts in the cross dispersion direction.  
\object{NGC~1068} provides perhaps the best example of this capability, since
it shows evidence for two fairly separated emission regions, which we
denote simply as ``primary'' and ``secondary.''  The
two cross-dispersion regions we use for the ``primary'' and ``secondary''
spots are shown in
Figs.~\ref{akinkhabwala-C2_fig:image} and \ref{akinkhabwala-C2_fig:3dimage}.  

\begin{figure*}[!ht]
  \begin{center}
    \begin{minipage}[t]{3.5in}
      \vspace{-.1in}\hspace{0in}
        \centerline{\psfig{file=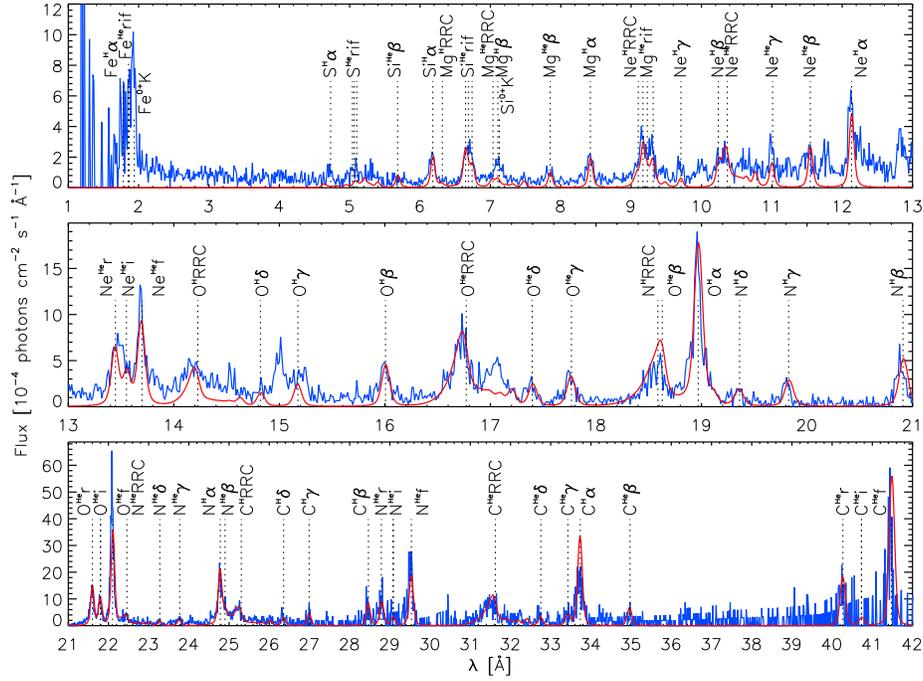, width=3.7in, angle=90}}
      \vspace{-.2in}
    \end{minipage}
  \end{center}
\caption{LETGS spectrum (m=-1/+1 orders) of the ``primary'' spot (blue)
and corresponding hydrogenic/heliumlike ion fit (red) using parameters 
listed in Table~\ref{akinkhabwala-C2_tab:table}.  The spectrum has been 
shifted to the \object{NGC~1068} rest frame ($z=0.00379$).  Model wavelengths 
are those expected for no excess velocity shifts.  We employ the same labelling
convention as in Fig.~\ref{akinkhabwala-C2_fig:full}.\label{akinkhabwala-C2_fig:primary}}
\end{figure*}

\begin{figure*}[!ht]
  \begin{center}
    \begin{minipage}[t]{3.5in}
      \vspace{-.1in}\hspace{0in}
        \centerline{\psfig{file=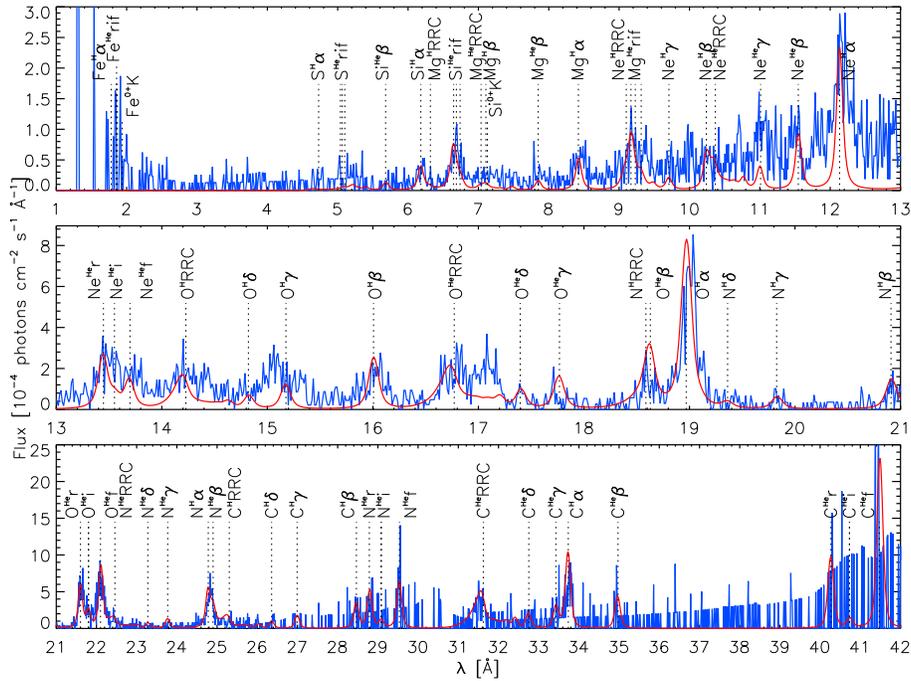, width=3.7in, angle=90}}
      \vspace{-.2in}
    \end{minipage}
  \end{center}
\caption{LETGS spectrum (m=-1/+1 orders) of the ``secondary'' spot (blue)
and corresponding hydrogenic/heliumlike ion fit (red) using parameters 
listed in Table~\ref{akinkhabwala-C2_tab:table}.  The spectrum has been shifted to the \object{NGC~1068} rest frame ($z=0.00379$).  Model wavelengths are those expected for no excess velocity shifts.  We employ the same labelling 
convention as in Fig.~\ref{akinkhabwala-C2_fig:full}.
Note the enhanced resonance 
line in the {\small{O}} and {\small{Ne}} heliumlike triplets.  
Several higher-order (np$\rightarrow$1s, $n>2$) lines also appear (e.~g., {\rm '$\mathrm{O^{He}\delta}$'} and {\rm '$\mathrm{N^{H}\gamma}$'}), providing 
clear evidence for photoexcitation rather than an additional hot, 
collisional gas component (see Fig.~\ref{akinkhabwala-C2_fig:PIPECIE}).\label{akinkhabwala-C2_fig:secondary}}
\end{figure*}

Spectra of the ``primary'' and ``secondary'' spots are given in 
Figs.~\ref{akinkhabwala-C2_fig:primary} and 
\ref{akinkhabwala-C2_fig:secondary}, respectively, along with their 
corresponding best fit models.  The fit parameters we obtain
are given in Table~\ref{akinkhabwala-C2_tab:table}.  The same $fL_X$ 
parameter works fairly well for all spectra, however a slightly different
radial velocity width $\sigma_v^{\mathrm{rad}}$ is preferred by the ``primary''
spectrum.  This illustrates the relative factor-of-two uncertainty in all fit 
parameters.  However, this uncertainty does not affect any of our 
conclusions.  All {\small LETGS} models have been convolved with the
specific zero-order dispersion profile.

The ``secondary'' spot, which is roughly coincident with the radio outflow 
lobes and the edge of the {\small{UV}} ionization cone, exhibits a noticeably
different 
spectrum from the ``primary'' spot.  The 
heliumlike resonance lines of Ne and O are relatively stronger compared to 
the forbidden lines than in the ``primary'' spot.
This enhancement might be due to an additional collisional component, 
perhaps outflow-shock-heated gas.  However, this is ruled out by the strength
of several higher-order-series lines and the normalization of the heliumlike 
intercomination/forbidden lines to the {\small{RRC}} (see 
Fig.~\ref{akinkhabwala-C2_fig:PIPECIE} for an explanation).  The observed  
enhancements are in fact simply due to a {\em lower} column density through 
the material (Fig.~\ref{akinkhabwala-C2_fig:opticaldepth}).

\section{Conclusions}

\begin{figure}[!ht]
  \begin{center}
    \begin{minipage}[t]{3.5in}
      \vspace{-.2in}\hspace{0in}
        \centerline{\psfig{file=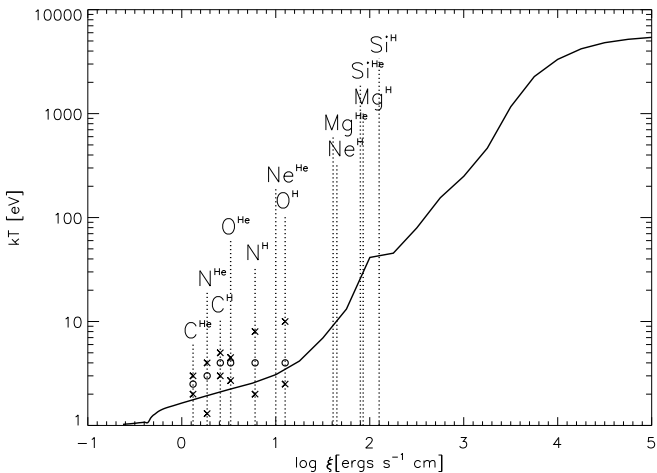, width=2.65in, angle=90}}
      \vspace{-.25in}
    \end{minipage}
  \end{center}
\caption{Recombination emissivity peak (for formation of the '{\rm H}'- or 
'{\rm He}'-like ionic species) calculated using {\small XSTAR} with 
incident power-law spectrum ($\Gamma=-1.7$, 13.6~eV$<E<$100~keV).  The 
{\small FWHM} of the emissivity for each ion is $\Delta\xi\sim1$.  The x's and 
o's give the confidence interval and best fit, respectively, for each ion 
temperature measured from the ion {\small RRC}.\label{akinkhabwala-C2_fig:ksi}}
\end{figure}

We have shown that the {\emph{XMM-Newton}} {\small{RGS}} spectrum of the 
soft-X-ray emission from \object{NGC~1068} is due entirely to recombinations 
and radiative cascade in an ionized gas cone, which is photoionized and 
photoexcited by the inferred nuclear continuum.  A simple model of a warm,
photoionized cone is capable of explaining all hydrogenic and heliumlike
ion spectra {\emph{in detail}}.  The values we infer for the radial ionic 
column densities
are similar to column densities observed in absorption (warm absorber)
in Seyfert~1 galaxies (\cite{akinkhabwala-C2:kaastra,akinkhabwala-C2:kaspi,akinkhabwala-C2:branduardi-raymont,akinkhabwala-C2:sakoiras}).  Since the
ionization cone in \object{NGC~1068} is spread over a region of $\sim$300~pc 
(\cite{akinkhabwala-C2:young}), this implies that generic warm absorbers 
have typical sizes of hundreds of parsec, rather than existing very close 
(e.~g., $<1$~pc) to the nucleus.
A subsequent spectrum obtained with {\emph{Chandra}} {\small{LETGS}} confirms 
the {\small{RGS}} results for the spectrum of the entire object, but also 
allows for spatially-resolved spectroscopy of the two bright spots resolved in 
the zero-order image.  The difference in spectra between these two spots is due
entirely to different radial ionic column densities, and not, for example, 
to an additional contribution from hot, outflow-shocked gas.
In Fig.~\ref{akinkhabwala-C2_fig:ksi}, we show that the temperatures we 
obtain for all spectra using well-measured, non-blended {\small RRC} are 
fairly consistent with (though slightly higher than) temperatures predicted 
from a self-consistent simulation using {\small XSTAR} 
(\cite*{akinkhabwala-C2:xstar}) of an optically-thin photoionized plasma.

\begin{figure}[!ht]
  \begin{center}
    \begin{minipage}[t]{3.5in}
      \vspace{0in}\hspace{.1in}
        \centerline{\psfig{file=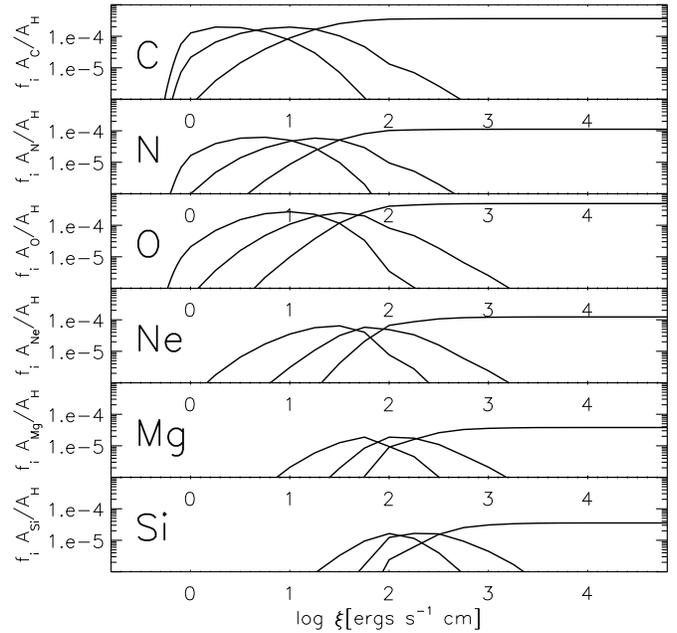, width=3.3in, angle=90}}
      \vspace{-.2in}
    \end{minipage}
  \end{center}
\caption{Fractional ionic abundances for the heliumlike (left), hydrogenic 
(middle), and bare (right) charge states for several ions as a function of 
ionization parameter $\xi=L_X/n_{\mathrm{e}} r^2$.\label{akinkhabwala-C2_fig:fractional}}
\end{figure}

The ionic column densities we infer allow us to probe the ionization parameter
distribution of the gas.  We show the predicted fractional ionic abundances as
a function of ionization parameter $\xi=L_X/n_{\mathrm{e}} r^2$ in 
Fig.~\ref{akinkhabwala-C2_fig:fractional}.  We point out the relatively equal 
column densities inferred for the hydrogenic and heliumlike species of each 
element in 
each spectrum in Table~\ref{akinkhabwala-C2_tab:table}.  No single ionization 
parameter is capable of reproducing this result.  Instead, a rather flat 
distribution in ionization parameter is necessary.
This could be obtained by assuming a spatially-stratified, single-density 
ionization cone (hence varying only $r$) or assuming an intrinsic density
distribution at each radius.  The striking overlap of {\small{O~III}} and 
soft-X-ray emission regions in \object{NGC~1068} 
(\cite{akinkhabwala-C2:young}) coupled with the presence of a similar range of 
ions in the ``primary'' and ``secondary'' spots, which are located at 
different distances from the nucleus, favor the latter interpretation.  The 
distribution in $\xi$ then is mostly due to a distribution in 
$n_{\mathrm{e}}$ of $f(n_{\mathrm{e}})\propto n_{\mathrm{e}}^{-1}$ over 
several orders of magnitude (roughly $n_{\mathrm{e}}\sim0.1$--$100$) at 
each radius (\cite{akinkhabwala-C2:rgs,akinkhabwala-C2:letgs}).

\begin{acknowledgements}
This work is based on observations obtained with {\em XMM-Newton}, an 
{\small ESA} science mission with instruments and contributions directly 
funded by {\small ESA} Member States and the {\small USA} ({\small NASA}).  
The Columbia University team is supported by {\small NASA}.  AK acknowledges 
additional support from an {\small NSF} Graduate Research Fellowship and 
{\small NASA GSRP} fellowship.  {\small MS} and {\small MFG} were partially 
supported by {\small NASA} through {\it   Chandra} Postdoctoral Fellowship 
Award Numbers 
{\small PF}01-20016 and {\small PF}01-10014, respectively, issued by the {\it Chandra} X-ray Observatory Center, which 
is operated by the Smithsonian Astrophysical Observatory for and behalf
of {\small NASA} under contract {\small NAS}8-39073.  {\small SRON} is 
supported by the Netherlands Foundation for Scientific Research 
({\small NWO}). Work at LLNL was performed under the auspices of the U. S. 
Department of Energy, Contract No. W-7405-Eng-48.
\end{acknowledgements}

\end{document}